\documentclass{iau} 
\usepackage{graphicx}

\title[RSGs in the Local Group] 
{The Red Supergiant Content \\ of the Local Group}

\author[Massey et al.]   
{Philip Massey$^1$,
Emily Levesque$^2$,
Kathryn Neugent$^1$,
Kate Evans$^{1,3}$,
Maria Drout$^4$,
\and Madeleine Beck$^5$}

\affiliation{$^1$ Lowell Observatory, 1400 W Mars Hill Road, Flagstaff, AZ 86001 and \\ Dept.\ Physics \& Astronomy, Northern Arizona University, Flagstaff, AZ 86011-6010 \\email: {\tt phil.massey@lowell.edu, kneugent@lowell.edu} \\[\affilskip]
$^2$Dept.\ Astronomy, University of Washington, Box 351580, Seattle, WA 98195 USA \\email: {\tt emsque@uw.edu}\\[\affilskip]
$^3$REU participant, 2015; California Institute of Technology, 1200 E, California Blvd, Pasadena, CA 91125 USA \\email: {\tt kevans@caltech.edu}\\[\affilskip]
$^4$Observatories of the Carnegie Institution for Science, 813 Santa Barbara St., Pasadena, CA 91101, USA \\email:{\tt mdrout@carnegiescience.edu}\\
[\affilskip]
$^5$REU participant, 2016; Wellesley College, 106 Central Street, Wellesley, MA 02481 USA\\email: {\tt mbeck4@wellesley.edu}}

\pubyear{2017}
\volume{329}  
\jname{The Lives and Death-throes of Massive Stars}
\editors{J.J.Eldridge,ed.}
\begin{document}

\maketitle

\begin{abstract}
We summarize here recent work in identifying and characterizing red supergiants (RSGs) in the galaxies of the Local Group.

\keywords{stars: early type,  supergiants, stars: Wolf-Rayet}
\end{abstract}

\firstsection 
\section{Introduction}

In the Olden Days, the term ``massive star" was synonymous with
``hot [massive] star," and topics at these conferences were restricted to
O-type stars and Wolf-Rayet (WR) stars.  Gradually massive star research has expanded to include
the red supergiants (RSGs).  Stars with initial masses $8-30M_\odot$ will evolve into RSGs, and these are the
progenitors of many core-collapse SNe. The amount of mass loss during the RSG phase 
will affect the subsequent evolution (if any) of these stars, with some evolving back to the blue side of
the H-R diagram (HRD), and possibly even becoming WRs (see, e.g., Meynet et al.\ 2015).

Studying RSGs in relation to other evolved massive stars provides an exacting test of our understanding
of massive star evolution, as these advanced stages act as a ``sort of magnifying glass, revealing relentlessly
the faults of calculations of earlier phases" (\cite[Kippenhahn \& Weigert 1991]{Mag}).  For instance, the relative number
of WRs and RSGs should be a very sensitive function of the initial metallicity of the host
galaxy, as first noted by \cite[Maeder et al.\ (1980)]{MaederAzz}.  In addition, the physical properties of RSGs
(effective temperatures, luminosities) provide an important check on the models.

Knowledge of the upper mass (luminosity) limit to becoming a RSG is also necessary to understand the relevance
of long-period binaries on the evolution of massive stars, and in particular, the formation of WRs.  Garmany et al.\ (1980) found that the close
binary frequency of O-type stars is 35-40\%, a number that has been confirmed by several recent studies (e.g., Sana et al.\ 2012, 2013).  High angular resolution surveys and long-term radial velocity studies have shown that if long period systems (as great as 10+ years) are included, the binary frequency is significantly higher, possibly 60\% or more (Caballero-Nieves et al.\ 2014, Aldoretta et al.\ 2015, Sana et al.\ 2012).  But, how much do these long period systems actually matter---do the components ever interact, or do they evolve as single stars within close sight of each other?  The answer to that depends upon how large the stars become.   Let's take Betegeuse as a ``typical" RSG; it has a radius (very roughly) of 5 AU.  Imagine it had an OB star companion (it doesn't): the stars would be in contact if the period were 2 years (assuming circular orbits) from Kepler's 3rd law, assuming masses of 15$M_\odot$+15$M_\odot$.   But, a 5 or 10 year period would be totally uninteresting in terms of mass loss or mass transfer, unless the orbit were highly eccentric.  And, above some limit (20$M_\odot$? 30$M_\odot$?) we don't expect stars to become RSGs---they will instead maintain relatively small radii as they evolve to the WR phase.  We can then see that it's important to measure what this limit is, so we can better understand how significant the binary channel actually is in the formation of WRs.  

In this short contribution, we'll explain how we find RSGs in nearby galaxies, how we characterize their properties, and what we plan to do next. Studying these stars in nearby galaxies allows us to explore stellar evolution as a function of metallicity.  
It's been recognized for 40 years that mass-loss plays a significant role in the evolution of massive stars.  As a result,
we expect there to be galaxy-to-galaxy differences in the relative numbers of WN- and WC-type WRs, and the relative numbers of WRs and RSGs.  This is because these stars will be born with different metallicities, and stellar wind mass-loss is driven through radiation pressure in highly ionized metal lines.  
Nearby star-forming galaxies can thus serve as our laboratories for carrying out such comparisons, as their metallicities range over a factor of $30\times$, from the low-metallicity Sextans A and B galaxies ($z=0.06\times$ solar), 
to the SMC and LMC ($z=0.27\times$ and $0.47 \times$ solar, respectively), to M33 (with a metallicity gradient going from $\sim$ solar in the center to SMC-like in the outer regions, to M31 ($1.6\times$ solar).

\section{Finding RSGs}

When we look at a nearby galaxy, we are looking through a sea of foreground stars in our own Milky Way.  The extent of this problem was first described by Massey (1998): years earlier, Humphreys \& Sandage (1980) had compared the distributions of  blue and red stars in M33.  It was a bit of a mystery why the former was clumpy, while the latter was smooth. Massey (1998) showed that there was significant contamination of the red star sample by foreground red dwarfs in our own Milky Way. 

 In the Magellanic Clouds the contamination of potential RSGs by foreground dwarfs can be nearly eliminated by the use of proper motions. In their study of LMC yellow and red supergiants, Neugent et al.\ (2012) showed that once they eliminated stars with proper motions greater than 15 mas~yr$^{-1}$ (about the limit of believability of cataloged proper motions pre-GAIA), the vast remainder of the red stars have radial velocities consistent with LMC membership.  RSGs in
 the LMC will have $V$ magnitudes of 9.5 to 12.5; foreground M0 dwarfs (say) in this magnitude range will be only 25-100 pc distant.

\begin{figure}[t]
\begin{center}
\includegraphics[width=2.6in]{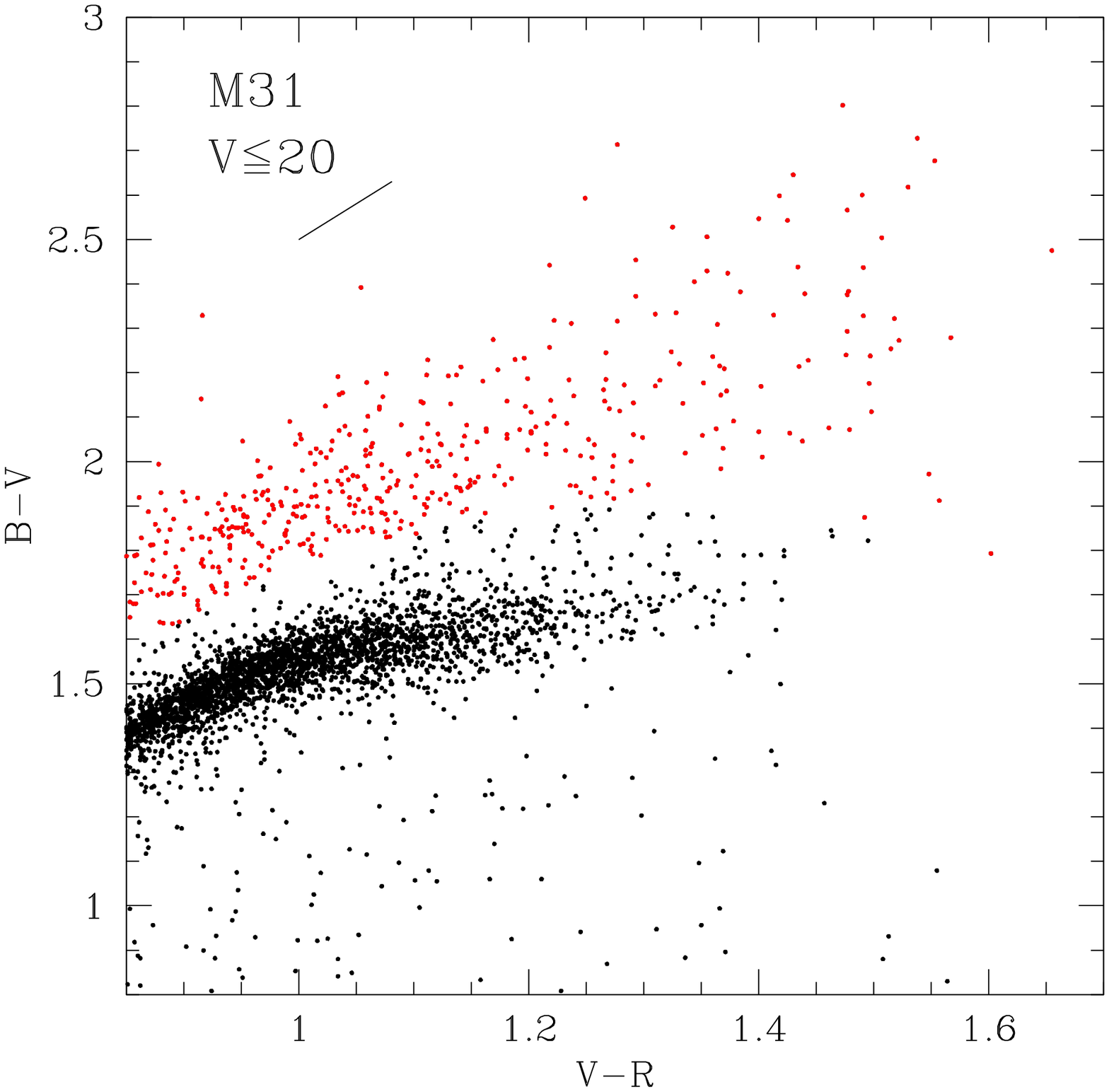} 
\includegraphics[width=2.6in]{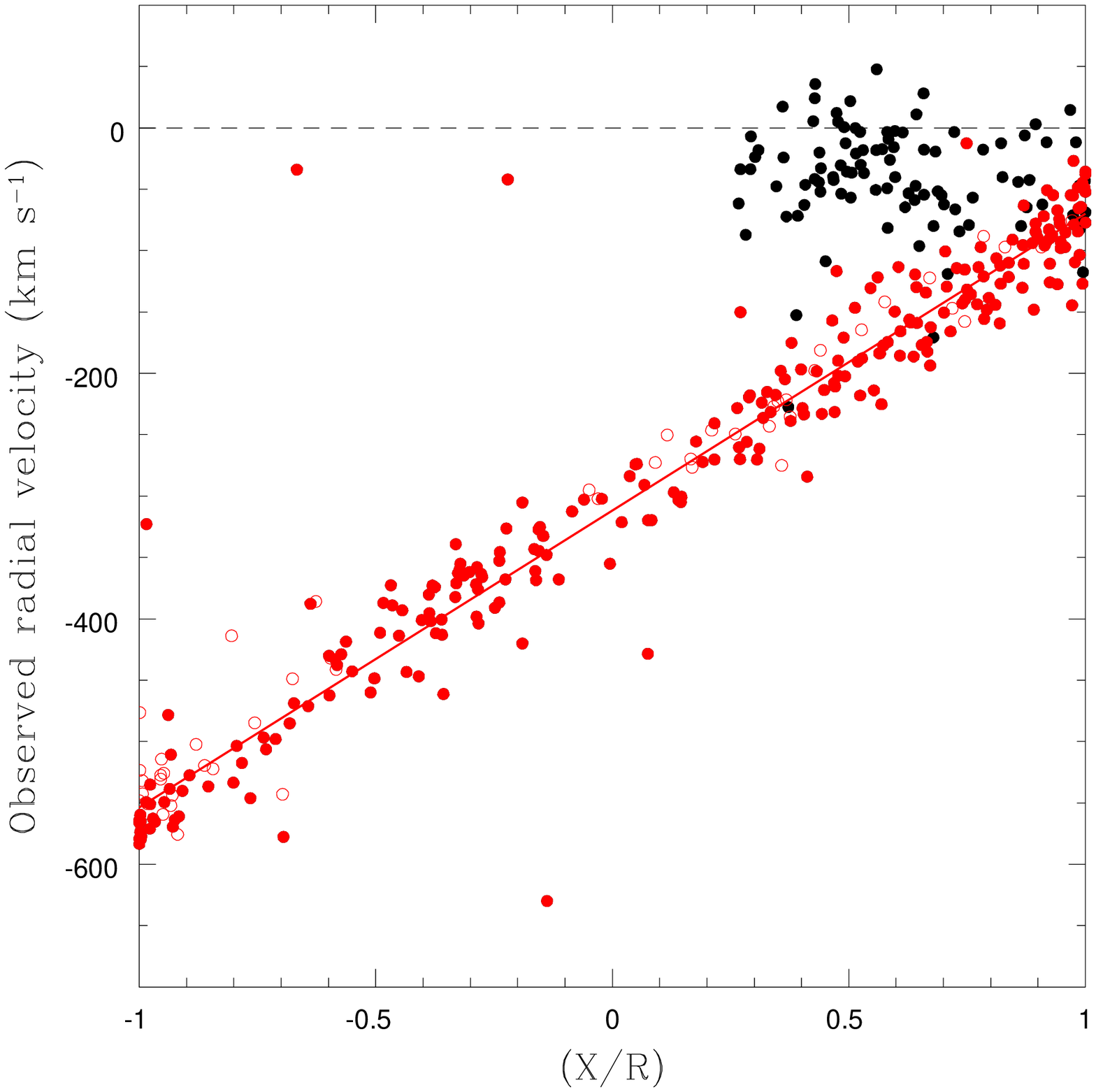}
\caption{Left: A two color diagram of red stars seen towards M31.  Stars expected to be RSGs on the basis of their {\it B-V} colors are colored red, and foreground stars, black.  Right:  The radial velocity of some of the sample stars. For the most part, stars we expect to be RSGs (red) follow the rotation curve of M31 (shown as a red line), while expected foreground stars (black) cluster around zero velocity.   $X/R$ denotes the position within M31.  From Massey \& Evans (2016).}
 \label{fig:kate}
\end{center}
\end{figure}

 For the more distant members of the Local Group, such as M31 and M33, proper motions are not useful: at $V$=17-20 (the magnitude range of their RSGs), foreground red dwarfs will be at larger distances (400-1500~pc).  Instead, Massey (1998) showed that one could use {\it B-V, V-R} two-color diagrams to separate supergiants from foreground stars.  At these colors, model atmospheres suggested that {\it B-V} would become primarily a surface gravity indicator due to line blanketing in the blue, while {\it V-R} remains a temperature indicator.  Radial velocities can be used to demonstrate the success of this, as shown in Fig.~\ref{fig:kate}.
 
\section{Physical Properties}
When our group began studying RSGs, we found that there was a bit of a problem: RSGs were considerably cooler and
more luminous than the evolutionary tracks could produce (see, e.g., Massey \& Olsen 2003).  In fact, the locations of these stars was in
the Hayashi forbidden zone (Hayashi \& Hoshi 1961), where stars were no longer in hydrostatic equilibrium.  Usually in such cases the problem lies with the theory, and indeed Maeder \& Meynet (1987) had demonstrated that how far over
the tracks extended to cooler temperatures depended upon how convection and mixing were treated.  Still, what if the
``observations" were wrong?  After all, we don't actually observe effective temperatures and bolometric luminosities; instead, we obtain photometry and spectral types and convert these to physical properties.

Levesque et al.\ (2005) revisited this issue using a new generation of model atmospheres, and found (to their relief) that the
new effective temperature scale was significantly warmer.  Indeed, when they plotted the location of their stars in the HRD,
there was near-perfect agreement with the evolutionary models, both in terms of the effective temperatures and upper luminosities.  Furthermore, extension of this work to lower and higher metallicity environments (Levesque et al. 2006, Levesque \& Massey 2012, Massey et al.\ 2009, Massey \& Evans 2016, Drout et al.\, 2012) have consistently shown that the change in the effective temperature scale tracks the expected metallicity-dependent change in the Hayashi limit.   This was dramatically shown  by Drout et al.\ (2012): as we move outwards in M33 to lower metallicities, the positions of RSGs shift to warmer temperatures.

\begin{figure}[h!]
\begin{center}
\includegraphics[width=2.4in]{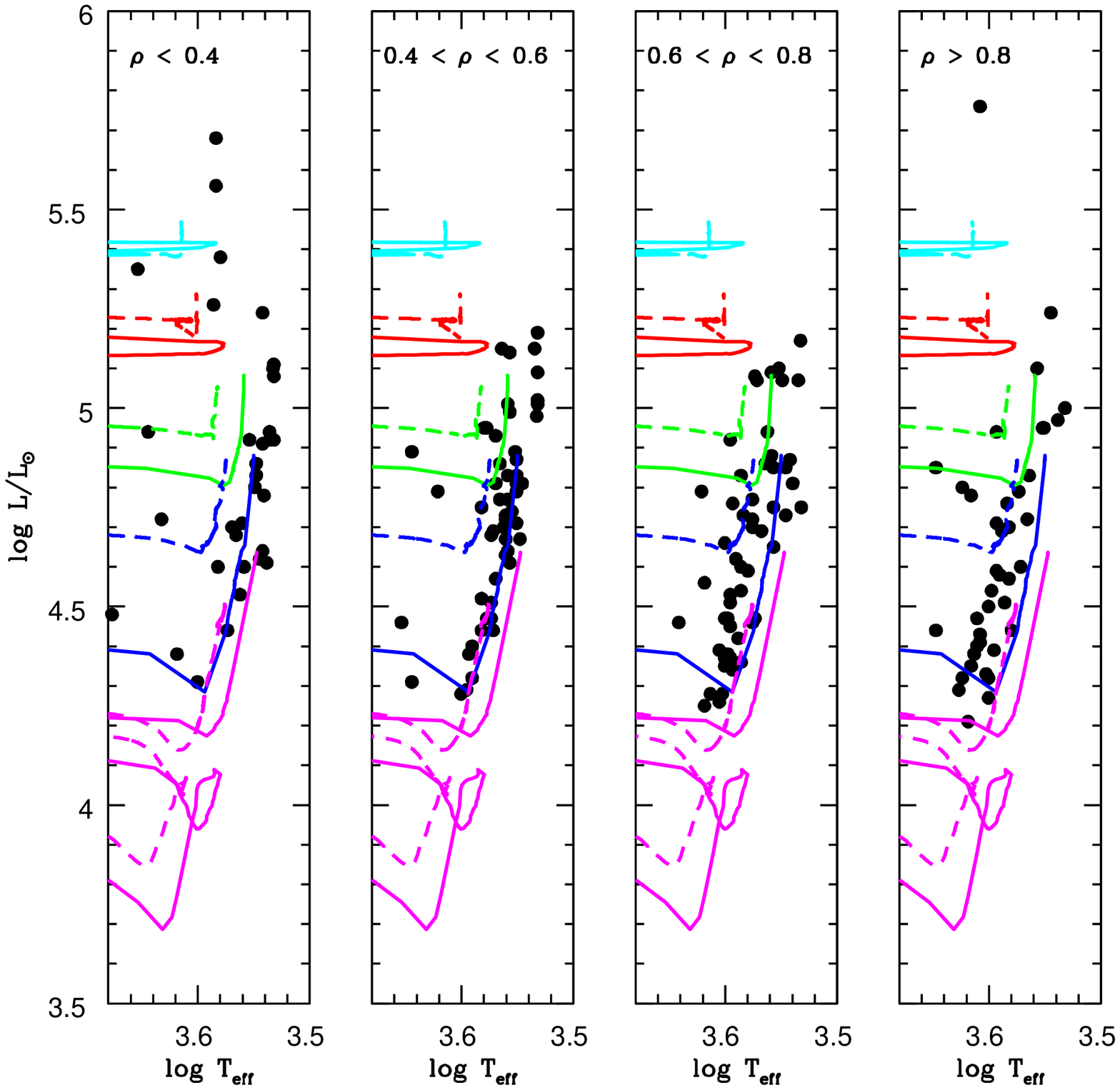} 
\includegraphics[width=2.4in]{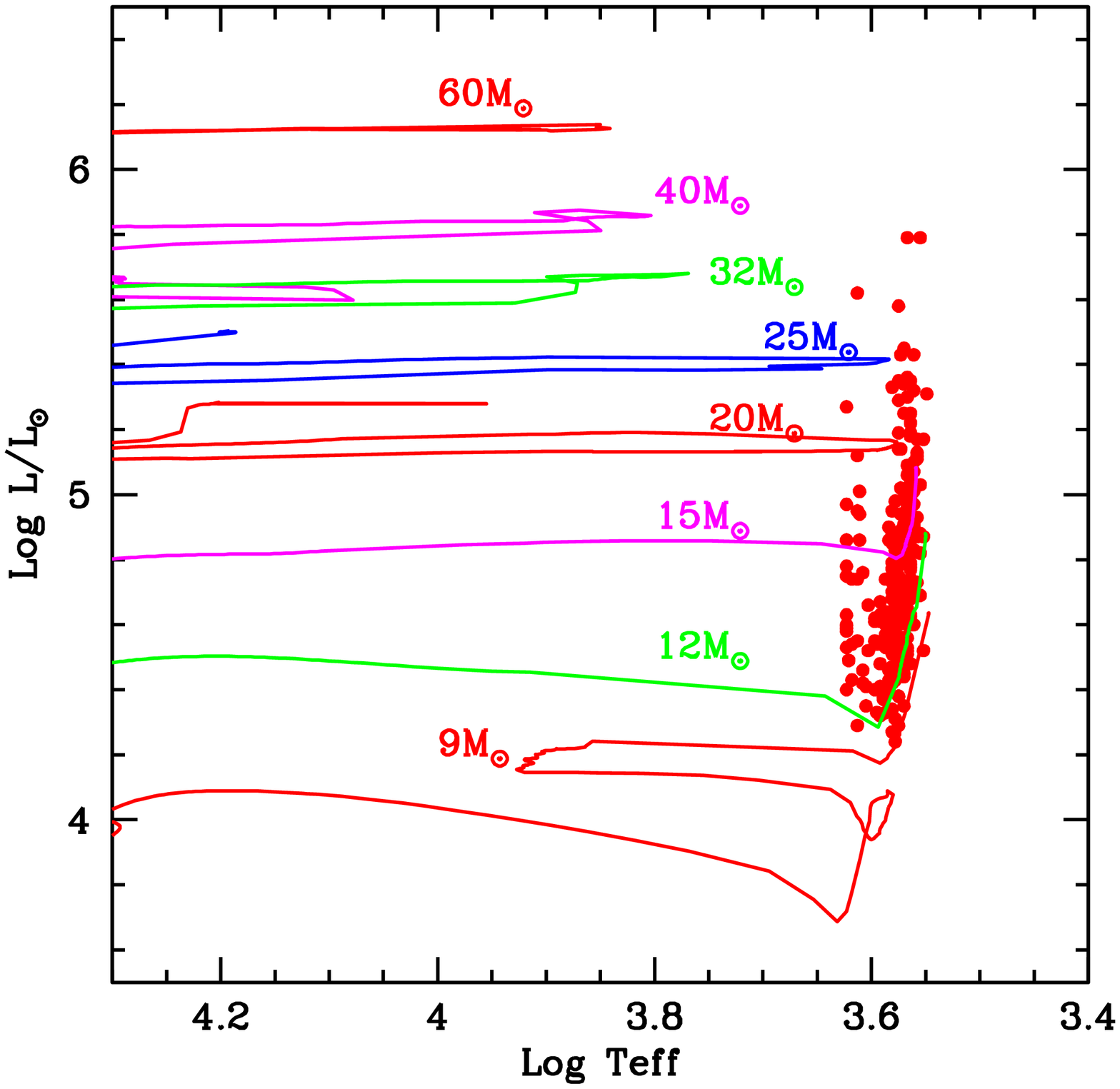}
\caption{Left: The change in effective temperatures of RSGs in M33 with galactocentric distance $\rho$.  At lower metallicities (greater $\rho$) RSGs are warmer.  This is in accord with the change of the Hayashi limit with metallicity. From Drout et al.\ (2012). Right: The HRD of RSGs in M31.  From Massey \& Evans (2016). }
 \label{fig:drout}
\end{center}
\end{figure}

When we are done with such studies, we can then compare the resulting HRD to that expected on the basis of the evolutionary tracks.  An example is shown in Fig.~\ref{fig:drout} (right) for RSGs in M31.  In this case the tracks plotted are for
solar metallicity (from Ekstr{\"o}m et al.\ 2012), which is a bit low for the metallicity of M31, but are the highest metallicity
modern tracks that are available at present. We see very good agreement for the most part.  The two highest
luminosity stars in the figure may or may not be members; see discussion in Massey \& Evans (2016). 

\section{Other Cool Things That We've Found}

Along the way, we have made some other interesting discoveries.  An inspection of Fig.~\ref{fig:kate} (right) reveals one star
($X/R\sim -0.15$) whose radial velocity is 300 km s$^{-1}$ discrepant with that expected from M31's rotation curve.  Was it just bad data?  We had four nice MMT spectra of the star, all yielding the same result.  Furthermore, the star is located about 4.5~kpc from M31's disk, about the distance a star could travel at 300 km s$^{-1}$ (assuming a transverse peculiar velocity similar to its excess radial velocity) in 10~Myr, about the lifetime of a massive star in the RSG stage.  Evans \& Massey (2015) identify this star as the first runaway RSG, and the first massive extragalactic runaway star. 

One of the consequences of the shifting of the Hayashi limit to warmer temperatures at lower metallicity is that
the spectral types of RSGs also shift: in the Milky Way, the average RSG is about an M2~I, in the LMC it's
about a M1~I, and in the SMC most RSGs are K5-7~I (Levesque et al.\ 2007). 
Massey et al.\ (2007) were therefore surprised to
find an M4~I star in the SMC, a star which was also one of the brightest RSGs, HV~11423.  The next year they reobserved
the star, only to find it was then a K0-1~I.  Did they just mess up?  Inspection of archival spectral showed that it been
caught in an M4.5-5 state on an ESO VLT spectrum taken years earlier, although no one had apparently found this remarkable at the time.  
Levesque et al.\ (2007) subsequently identified seven LMC and four additional SMC RSGs that were also later than expected
for spectral types in their host galaxies; all of these (including HV~11423) show large $V$-band variability, and change their
effective temperatures by 3-4\% on the timescales of months.  At their coolest, these stars are well into the Hayashi forbidden zone.

We have no understanding of what these stars are from an evolutionary point of view, the finding was intriguing enough for Anna \.{Z}ytkow to contact us and suggest that perhaps these stars might be the long-sought after Thorne-\.{Z}ytkow Objects (T\.{Z}Os).  T\.{Z}Os were first proposed  Thorne \& \.{Z}ytkow (1975, 1977); these are RSGs that
have neutron star cores and hence a variety of very interesting nuclear reactions.  Such objects might be the
primary source of the proton-rich nuclei in the Universe, solving a long-standing mystery (Cannon 1993).  The only problem was that no viable T\.{Z}O candidate had ever been identified.  None of the Levesque-Massey variables prove to show the
weird elemental enrichment expected of T\.{Z}O, but another RSG, selected for being too cool, did, HV~2112 (Levesque et al.\ 
2014)\footnote{Maccarone \& de Mink (2016) suggest that HV 2112 is not a member of the SMC, but this was based on an unfortunate reliance on the SPM4 proper motion catalog, which is known to contain significant errors; other, more reliable catalogs show null proper motions, and the claim was quickly refuted by Worley et al.\ (2016).  GAIA will of course answer this definitively.}.

\section{Where Do We Go From Here?}
We have provided a quick snap-shot here of where we stand in the identification of RSGs in the Local Group, and what we have learned from these stars.  We are working on characterizing complete samples of RSGs in the Magellanic Clouds, Sextans A and B, and other Local Group galaxies.  Hopefully by the time of the next massive stars conference we will be able to
report on clean relative numbers of RSGs and WRs throughout the galaxies of the Local Group, and compare these to both the Geneva single-star and BPASS (Eldridge \& Stanway 2016) binary evolutionary models.


\begin{thebibliography}{}

\bibitem[]{}Aldoretta, E. J., Caballero-Nieves, S.. M., Gies, D. R. et al.\ 2015, \textit{AJ}, 149, 26

\bibitem[]{}Caballero-Nieves, S. M., Nelan, E. P., Gies, D. R. et al. 2014, \textit{AJ}, 147, 40

\bibitem[]{}Cannon, R. C. 1993, \textit{MNRAS}, 263, 817

\bibitem[]{}Drout, M. R., Massey, P., \& Meynet, G. 2012, \textit{ApJ}, 750, 97

\bibitem[]{}Ekstr{\"o}m, S., Georgy, C., Eggenberger, P. et al.\ 2012, \textit{A\&A}, 537, A146

\bibitem[]{}Eldridge, J. J., \& Stanway, E. R. 2016, \textit{MNRAS}, 462, 3302

\bibitem[]{}Evans, K. A., \& Massey, P. 2015, \textit{AJ}, 150, 149

\bibitem[]{}Garmany, C. D., Conti, P. S., \& Massey, P. 1980, \textit{ApJ}, 242, 1063

\bibitem[]{}Hayashi, C., \& Hoshi, R. 1961, \textit{PASJ}, 13, 442

\bibitem[]{}Humphreys, R. M., \& Sandage, A.1980, \textit{ApJS}, 44, 319

\bibitem[Kippenhahn \& Weigert (1991)]{Mag}
{Kippenhahn, R., \& Weigert, A.} 1990, \textit{Stellar Structure and Evolution} (Berlin: Springer-Verlag), 192

\bibitem[]{}Levesque, E. M., \& Massey, P. 2012, \textit{AJ}, 144, 2

\bibitem[]{}Levesque, E. M., Massey, P., Olsen, K. A. G., Plez, B. 2007, \textit{ApJ} 667, 202

\bibitem[]{}Levesque, E. M., Massey, P., Olsen, K. A. G., et al.\  2005, \textit{ApJ}, 628, 973

\bibitem[]{}Levesque, E. M., Massey, P., Olsen, K. A. G., et al.\  A. 2006, \textit{ApJ}, 645, 1102

\bibitem[]{}Levesque, E. M., Massey, P., \.{Z}ytkow, A. N., \& Morrell, N. 2014, \textit{MNRAS}, 443, L94

\bibitem[]{}Maccarone, T. J. \& de Mink, S. E. 2016, \textit{MNRAS}, 458, 1

\bibitem[Maeder et al.\ (1980)]{MaederAzz}
{Maeder, A., Lequeux, J., \& Azzopardi, M.} 1980, \textit{A\&A}, 90, L17

\bibitem[]{}Maeder, A., \& Meynet, G. 1987, \textit{A\&A}, 182, 243

\bibitem[]{}Massey, P. 1998, \textit{ApJ}, 501,153

\bibitem[]{}Massey, P., \& Evans, K. A. 2016, \textit{ApJ}, 826, 224

\bibitem[]{}Massey, P., Levesque, E. M., Olsen, K. A. G., Plez, B., \& Skiff, B. A. 2007, \textit{ApJ}, 660, 301

\bibitem[]{}Massey, P., \& Olsen, K. A. G. 2003, \textit{AJ}, 126, 2867

\bibitem[]{}Massey, P., Silva, D. R., Levesque, E. M. 2009, \textit{ApJ}, 703, 420

\bibitem[]{}
Meynet, G., Chomienne, V., Ekstr\"{o}m, S. et al.\  2015, \textit{A\&A}, 575, A60

\bibitem[]{}Neugent, K. F., Masey, P., Skiff, B., \& Meynet, G. 2012, \textit{ApJ}, 749, 177

\bibitem[]{}Sana, H., de Mink, S. E., de Koter, A. et al.\ 2012, \textit{Sci}, 337, 444

\bibitem[Sana et al.\ (2013)]{Sana13}
{Sana, H, de Koter A., de Mink, S. E. et al.} 2013, \textit{A\&A}, 550, A107

\bibitem[]{}Thorne, K. S., \& \.{Z}ytkow, A. N. 1975, \textit{ApJ}, 199, L19

\bibitem[]{}Thorne, K. S., \& \.{Z}ytkow, A. N. 1977, \textit{ApJ}, 212, 832

\bibitem[]{}Worley, C. C., Irwin, M. J., Tout, C. A. et al.\ 2016,
\textit{MNRAS}, 459, L31.

\end{thebibliography}
\end{document}